\documentclass[seceq]{ptptex}

\usepackage{wrapft}

\def\any{{}^{\forall}}

\def\itmb{\begin{enumerate}}
\def\itme{\end{enumerate}}
\def\eqnb{\begin{equation}}
\def\eqne{\end{equation}}
\def\eqab{\begin{eqnarray}}
\def\eqae{\end{eqnarray}}
\def\eqsb{\begin{eqnarray*}}
\def\eqse{\end{eqnarray*}}
\def\dis{\displaystyle}

\def\scsc{\scriptscriptstyle}

\def\jm5{{j_{\scsc{\mu}}^{\scsc{5}}}}

\def\subeqb{\begin{subequations}}
\def\subeqe{\end{subequations}}

\def\any{{^{\scsc{\forall}}}\!}
\notypesetlogo  

\title{
Horizon function in Landau gauge QCD revisited\\
--Free boundary case from viewpoint of network QCD--}
\author{
Hideo \textsc{Nakajima}\footnote{E-mail: nakajima@is.utsunomiya-u.ac.jp}
}

\inst{
Department of Information Systems Science, Utsunomiya University,\\
Utsunomiya 321-8585, Japan
}

\date{}
\abst{
In Gribov-Zwanziger scenario\cite{rf:11} of color confinement, Zwanziger proposed two types of horizon function in lattice Landau gauge, the first 
one\cite{rf:1} and the second one of which foundation was discussed in detail\cite{rf:2}. 
The second type horizon function is focussed on in the present study. Its derivation and the horizon condition are briefly reviewed along the line of Ref.~\citen{rf:2}, and it is also reviewed that this horizon condition and Kugo-Ojima color confinement criterion\cite{rf:3} coincide in the continuum limit.\cite{rf:6,rf:7} In case of free boundary condition in contrast to periodic boundary condition, it was shown that the horizon condition holds for each gauge fixed configuration in Landau gauge.\cite{rf:5}. It is clarified that this fact can be derived from some identity relation which holds on an arbitrary network of links in Landau gauge. 
Thus the origin of the pointwise validity of horizon condition is highlighted.  
}

\begin{document}
\maketitle

\vspace*{2mm}
\section{Notations and generalities}
We intend to define Landau gauge $SU(N)$ QCD on general networks, and to
analyze standard lattice Landau gauge QCD from general point of view.
So we have sites $x$ and links $\ell=(x,x')$;pairs of sites 
in general networks. 
To each link $\ell$ is assigned its intrinsic (positive) direction $e_\mu$ 
as convention, say $e_\mu=x'-x$, for specification of 
basic link variable $U_{e_\mu}$ 
as a parallel transport $U_{x,x'}$ in the direction of $e_\mu$, and of 
link-field $A_\mu$ (as "network current"), $A_{x,\mu}$ 
denote a value on the link $\ell=(x,x+e_\mu)$ called as $e_\mu$-component. 
Now links $\ell$ with $e_{\mu}$ are 
considered as vectors $xx'(=\overrightarrow {xx'})$. 
We use notations $U_{xx'}=U_{\ell}=U_{x,e_{\mu}}=U_{x,\mu}$ 
interchangeably, where $x'-x=e_\mu$. Let $\ell(x)$ denote a
set of links,
\eqnb
\ell(x)=\{\ell|\ \ell=xx'=e_\mu\ {\rm or}\ \ell=x'x=e_\mu\}
\eqne
and let $\ell_+(x)$ be defined as a set of {\em positive link at} $x$,
\eqnb
\ell_+(x)=\{\ell|\ \ell=xx'=e_\mu\}
\eqne
and similarly $\ell_-(x)$ a set of {\em negative link at} $x$,
\eqnb
\ell_-(x)=\{\ell|\ \ell=x'x=e_\mu\}.
\eqne
It is to be noted that number of elements of $\ell_+(x)$ is not necessarily equal to that of $\ell_-(x)$ in general networks in contrast to periodic regular lattice or infinite regular lattice.

Thus gauge transformation by $g\in G$ is 
written for $\ell\in \ell_+(x)$ as 
\eqnb
U^g_{\ell}={g_x}^\dag U_{\ell} g_{x+e_\mu},
\eqne
and for $\ell\in \ell_-(x)$ as 
\eqnb
U^g_{\ell}={g_{x-e_{\mu}}}^\dag U_{\ell} g_{x}.
\eqne
We denote normalized antihermitian 
matrices $\lambda_a$ as Lie algebra basis as
\eqnb
[\lambda_a,\lambda_b]=f_{abc}\lambda_c
\eqne
and
\eqnb
(\lambda_a|\lambda_b)={\rm tr}(\lambda^{\dag}_a\lambda_b)=\delta_{ab}.
\eqne
All fields of the adjoint representation are often denoted as antihermitian
fields in use of the above basis.

We use bracket notation for suitable innerproducts for scalar fields
 (site function) and vector fields (link function), respectively, as
\eqnb
\langle \psi|\phi\rangle=\sum_x {\rm tr} ({\psi_x}^{\dag}\phi_x),
\eqne
\eqnb
\langle A_\mu|B_\mu\rangle=\sum_{x,\mu} {\rm tr} 
({A_{x,\mu}}^{\dag}B_{x,\mu}),
\eqne
where we use a simple notation for link-functions, e.g., $A_\mu$, 
and $\dis{\sum_{x,\mu}}$ implies summation over $x$ and $\ell_+(x)$, or over $x$ and $\ell_-(x)$,  which is equivalent to summation over all links $\dis{\sum_{\ell}}$.
From site-fields (scalars) $\phi$, 
two kinds of link-fields (vectors) are defined such that
\eqnb
\partial_\mu \phi=\phi_{x+e_\mu}-\phi_x,
\eqne
according to the associated positive direction $e_\mu$ of the link 
$(x,x+e_\mu)$, and 
\eqnb
{\bar \phi}^\mu=(\phi_{x+e_\mu}+\phi_x)/2,
\eqne
where we use hereafter abbreviated notations $_{x+\mu}$ for 
$_{x+e_\mu}$.\\

The following relations hold on any networks as well as on regular 
lattice.
\eqnb
\langle A_{\mu}|\partial_\mu \phi \rangle
=\langle -\partial_\mu A_{\mu}| \phi \rangle,
\eqne
where the divergence of link-field is defined as
\eqnb
(\partial_\mu A_\mu)_x=\sum_{\mu\in \ell_+(x)} A_{x,\mu}
-\sum_{\mu\in \ell_-(x)} A_{x-\mu,\mu}.
\label{DIVLESS}
\eqne
Adjoint of commutator reads as
\eqnb
\langle A_{\mu}| [B_\mu C_\mu] \rangle
=\langle -[B_\mu A_{\mu}]| C_\mu \rangle.
\eqne
Adjoint of ${\bar \ast}^\mu$ reads as
\eqnb
\langle A_{\mu}|{\bar \phi}^\mu \rangle
=\langle {\bar A_{\mu}}^\mu| \phi \rangle.
\eqne
where a site function is defined as
\eqnb
({\bar A_{\mu}}^\mu)_x={1\over 2}\left \{\sum_{\mu\in \ell_+(x)} A_{x,\mu}
+\sum_{\mu\in \ell_-(x)} A_{x-\mu,\mu}
\right \}.
\eqne
\section{Definitions of gauge field and covariant derivative}
There are two possible options of $A_{\mu}(U)$,\\
$U$-linear definition;
\eqnb
A_{x,\mu}=(U_{x,\mu}-U_{x,\mu}^{\dag})/2|_{\rm traceless\ part},
\eqne
$\log\ U$ definition\cite{rf:NF}; 
\eqnb
U_{x,\mu}=e^{A_{x,\mu}}.
\eqne
Next we define a covariant derivative $D_\mu(U)$ which appears under an 
infinitesimal gauge transformation $e^\varepsilon$ as $\delta A_{x,\mu}=D_\mu(U)\varepsilon$;
\eqnb
D_{\mu}(U)\phi=G(U_\mu)\partial_\mu \phi+[A_\mu,\ {\overline{\phi}}^\mu],
\label{CVDR1}
\eqne
where the operation $G(U_\mu)$ on an antihermitian link variable, $B_\mu$, is given by,\\ 
$U$-linear definition;
\eqnb
G(U_\mu)B_\mu=
\dis{1\over 2}\left .\left\{ \dis{U_\mu+U_\mu^\dag\over 2},
{B_\mu}\right\}\right |_{\rm traceless\ part},
\eqne
$\log\ U$ definition; 
\eqnb
G(U_\mu)B_\mu\equiv S({\mathcal A}_{\mu})B_{\mu}
=\{({\mathcal A}_{\mu}/2)/{\rm th}({\mathcal A}_{\mu}/2)\}B_{\mu}
\eqne 
with 
\eqnb
{\mathcal  A}_{\mu}=adj_{A_{\mu}}=[A_{\mu},\cdot].
\eqne
It is to be noted that in both definitions $G(U_\mu)_{ab}=G(U_\mu)_{ba}$ from
\eqnb
G(U_\mu)_{ab}={\rm tr}\left (\lambda_a^{\dag}
\dis{1\over 2}\left\{ \dis{U_\mu+U_\mu^\dag\over 2},
\lambda_b\right\}\right )=G(U_\mu)_{ba}.
\eqne
and
\eqnb
{({\mathcal  A}_{\mu})^2}_{ab}={\rm tr}\left (\lambda_a^{\dag}
[A_\mu[A_\mu,\lambda_b]]\right )=
{\rm tr}\left ([A_\mu,\lambda_a][A_\mu,\lambda_b]]\right )
={({\mathcal  A}_{\mu})^2}_{ba},
\eqne
respectively.

Adjoint of the covariant derivative is defined as
\eqnb
\langle B_{\mu}|D_\mu(U) \phi \rangle
=\langle -D_\mu (U)B_{\mu}| \phi \rangle,
\eqne
where a site function $D_{\mu}(U)B_\mu$, with $\mu$ summation understood, is given as
\eqnb
D_{\mu}(U)B_\mu=\partial_\mu (G(U_\mu) B_\mu)+\overline{[A_\mu,B_\mu]}^\mu.
\label{CVDR2}
\eqne
\section{Optimization function and the Landau gauge}
The Landau gauge $\partial A=0$ can be characterized\cite{rf:MN} 
such that 
\eqnb
\delta F_U(g)=0\ \ {\rm for\ }~\any\delta g,
\eqne
in use of the optimizing functions $F_U(g)$ for each option of $A_\mu(U)$ as\\
$U$-linear definition; 
\eqnb
F_U(g)=\sum_{x,\mu}{\rm tr}\left \{2-\left (U^g_{x,\mu}+{U^g_{x,\mu}}^\dag\right)
\right \},
\eqne
$\log U$ definition;
\eqnb
F_U(g)=\sum_{x,\mu}{\rm tr} \left({{A^g}_{x,\mu}}^{\dag}\ A^g_{x,\mu}\right)
\equiv \langle A^g_\mu|A^g_\mu \rangle.
\eqne
It is seen that in case of infinitesimal 
gauge transformations $g^{-1}\delta g=\varepsilon$, 
variation $\delta F_U(g)$ is given in either definition as
\eqnb
\delta F_U(g)=2\langle A^g_\mu|\partial_\mu \varepsilon\rangle,
\eqne
that is, by putting $g=g(t)$, and $g^{-1}g_t=\omega_t$,
\eqnb
{d\over dt}F_U(g(t))=2\langle A^g_\mu|\partial_\mu \omega_t \rangle.
\eqne
It holds on the arbitrary networks that 
\eqnb
\delta F_U(g)=-2\langle \partial_\mu A^g_\mu|\varepsilon\rangle,
\eqne
or
\eqnb
{d\over dt}F_U(g(t))=-2\langle \partial_\mu A^g_\mu| \omega_t \rangle,
\eqne
which verifies the statement that a stationarity point of the optimizing
function on a gauge orbit yields the Landau gauge.
If we proceed further to a higher derivative in general, then
we obtain that
\eqnb
{d^2 F_U(g(t))\over dt^2}
=-2\langle \partial_\mu D_\mu(U^g)\omega_t| \omega_t \rangle
-2\langle \partial_\mu A^g_\mu| \omega_{tt} \rangle.
\eqne
Thus if $U^{g(0)}_\mu=U_\mu$, then the variation of the optimizing function, 
$\Delta F=F_U(e^\varepsilon)-F_U(1)$, is given up to the second order as
\eqnb
\Delta F=-2\langle \partial_\mu A_\mu|\varepsilon\rangle
+\langle \varepsilon |-\partial_\mu D_\mu(U) |\varepsilon \rangle,
\eqne
where it is put that $\varepsilon=\eta \Delta t$ with any site function $\eta$ constant with respect to 
$t$, i.e., $\omega_{tt}=0$ may be assumed in this case.\\

Let us denote after Zwanziger, the Landau gauge space as
\begin{equation}
\Gamma\equiv\{U|\partial A=0\},
\end{equation}
and the Gribov region as
\begin{equation}
\Omega\equiv \{U|M(U)\ge 0,\ U\in \Gamma\},
\end{equation}
where $M(U)$ is a Faddeev-Popov operator, $M=-\partial D(U)$, 
detailed properties of which will be investigated below.

Now one can define a 
{\bf fundamental modular region $\Lambda$} as a set of global minima of
the optimizing function.
\eqnb
\Lambda=\{ U|\ F_U(1)\le F_U(g)\ {\rm for\ }^\forall g\}.
\eqne

It holds the following inclusion as
\eqnb
\Lambda \subset \Omega \subset \Gamma.
\eqne 

So far, {\em all notions and formula are valid on any networks, i.e.,  in
any topology, and/or boundary conditions}.\\

For a while from now on, we assume $d$-dimensional regular lattice with $L$-periodic
boundary conditions, and a set of link-field $U$ on this lattice 
is denoted as $\Pi_L$, and a set of gauge transformation on this
lattice, $G_L$. We define {\bf the $N$th partial core of the 
fundamental region} as
\begin{equation}
\Lambda_L^{N}\equiv \{U|\ U\in \Pi_L\ {\rm and}\ 
F_U(1)\le F_U(g),\ {\rm for\ }^\forall g\in G_{NL}\}.
\end{equation}
From $G_L\subset G_{NL}$, inclusion $\Lambda_{L}^N \subset \Lambda_L$ is easily
understood, and one may write 
\begin{equation}
\Lambda_L^{N}=\Lambda_{LN}\cap \Pi_L ,
\end{equation}
and 
if $N$ is chosen to be a power of 2, say, then the partial cores are nested,
\begin{equation}
\Lambda_L^{N'}\subset \Lambda_L^N\subset \Lambda_L\ \ {\rm for}
\ N'=2^{M'}>N=2^M.
\end{equation}
The {\bf core region} is defined by the limiting set,
\begin{equation}
\Xi_L\equiv\Lambda_L^\infty=\lim_{N\to\infty}\Lambda_L^N=\Lambda_{\infty}\cap \Pi_L.
\end{equation}

In connection to the core region, the following theorem holds.\\

{\bf Theorem 1} \\
Let $U\in \Pi_L$, where $\Pi_L$ is a set of $L$-periodic configurations.
The gauge transformation $g$ which brings $U$ to $\Lambda_{NL}$ has
a form such that 
\eqnb
g=he^{\theta x}
\eqne
where 
\eqnb
h\in G_{L},\ \ [\theta_\mu,\theta_\nu]=0\ \ {\rm and}\ \ 
e^{\theta_\mu LN}=1.\\
\eqne

Proof\\
In the following, $U$ is commonly used as denoting a configuration 
either being $L$-periodic or being $NL$-periodic.
Let $U^g\in \Lambda_{NL}$ where $g(x)\in G_{NL}$.
Since the optimizing function $F$ is an extensive quantity,
it holds that the shifted configuration in any direction 
by lattice unit have the same value of $F$. Particularly,
the shifted configuration of $U^g$ in the negative $\mu$-direction
by $L$ units is also a gauge transform of $U$ into $\Lambda_{NL}$,
and it can differ from $U^g$ only by constant gauge transformaton,
say, $g_\mu$. Now since the shifted configuration $U(x+L\mu)$ is 
identical to $U(x)$ itself, 
it follows that $g$ has a structure such that the $L$-unit
shifted $g$, i.e., $g(x+L\mu)$ in the negative $\mu$-direction is given by
$g(x)g_\mu$. It is obvious that $[g_\mu, g_\nu]=0$, 
since translations in any directions commute each other, and 
$g_\mu^N=1$ from $NL$-periodicity of $g$.
We may write $g_\mu=e^{\theta_\mu L}$ where $\theta_\mu$'s belong to
the same Cartan subalgebra, and $e^{\theta_\mu NL}=1$.
Let $h(x)=g(x)e^{-\theta x}$, then it follows that
\eqnb
h(x+L\mu)=g(x+L\mu)e^{-\theta_\mu L}e^{-\theta x}=g(x)g_\mu g_\mu^{-1}e^{-\theta x}
=h(x),
\eqne
and thus $g=he^{\theta x}$ and $h\in G_L$.
q.e.d.\\

Remark 2\\
Let $U\in \Lambda_L^N$. It holds that
\eqnb
F_U(1)\le F_U(g)\ \ {\rm for\ }^\forall g=he^{\theta x},\ \ (h\in G_L)
\eqne 
where $\theta_\mu=(M/NL)\eta_\mu$ with the nonzero smallest 
elements $\eta_\mu$ of the Cartan subalgebra such that 
$e^{\eta_\mu} =1$, and $M$ is an arbitrary integer. It is to be
noted here that $U^g \in \Pi_{NL}$, but $U^g \notin \Pi_{L}$ in general.\\

Proof is selfevident from the definition of $\Lambda_L^N$. 
\\

Remark 3\\
Let $U\in \Xi_L$. It holds that
\eqnb
F_U(1)\le F_U(g)\ \ {\rm for\ }^\forall g=he^{\theta x},\ \ (h\in G_L)
\eqne 
where $\theta_\mu=t\eta_\mu$ with the nonzero smallest 
elements $\eta_\mu$ of the Cartan subalgebra such that 
$e^{\eta_\mu} =1$, and $t$ is an arbitrary real. Similarly to Remark 2,
it is noted that $U^g \in \Pi_{\infty L}$ in general.
\\

Proof is selfevident from the definition of $\Xi_L$. 
\\

Now we investigate the behavior of the optimizing functions $F_U(g)$
under the gauge transformation relaxed so as to include
the gauge transformation of the Bloch wave type.

Let $g$ be as $g=he^{\theta x}=e^\omega e^{\theta x}\equiv e^\xi$,
and let us consider that $\theta_\mu$ and $\omega$ are some
functions of $t$ such that
\eqnb
\theta_\mu=t \eta_\mu\ \ {\rm and\ \ }\omega=\omega(t)
\eqne
where $\eta_\mu$'s are suitably normalized constant 
elements of Cartan subalgebra,
and $\omega(t)$ is a $L$-periodic scalar field with $\omega(0)=0$.
Then putting 
\eqnb
h_t\equiv\dis{d h\over dt}\equiv h \omega_t
\ \ {\rm and\ \ }
g_t\equiv\dis{d g\over dt}\equiv g\xi_t,
\eqne
we obtain
\eqnb
\xi_t= e^{-\theta x}\omega_te^{\theta x}+\eta x\equiv \omega'_t+\eta x,
\label{XIT}
\eqne
where it is to be noted that neither ${d\omega\over dt}=\omega_t$ nor 
${d\xi\over dt}=\xi_t$ hold in general. Another point to be emphasized here
is that $U^g$ is considered as gauge transform of $U\in \Pi_L\subset 
\Pi_{\infty L}$ and is not of $L$-periodicity in general.
It follows from general derivation before that
\eqnb
{d\over dt}F_U(g(t))=2\langle A^g_\mu|\partial_\mu \xi_t \rangle.
\eqne
From (\ref{XIT}), we have
\eqnb
{d\over dt}F_U(g(t))=2(\langle A^g_\mu|\partial_\mu \omega'_t \rangle
+\langle A^g_\mu|\eta_\mu \rangle
),
\label{FUT0}
\eqne
and
\eqnb
{d\over dt}F_U(g(t))=2(-\langle \partial_\mu A^g_\mu| \omega'_t \rangle
+\langle A^g_\mu|\eta_\mu \rangle
).
\label{FUT}
\eqne
The above equation (\ref{FUT}) is trivial as one on $\infty L$-periodic 
lattice, but it has more implication than that. 
Although the periodicity of $\omega'_t$ and $A^g_\mu$ can not be 
demonstrated to be $L$, actual contribution from each link in (\ref{FUT0})
appears $L$-periodic, and the same holds in (\ref{FUT}), and thus
the derivation of (\ref{FUT}) can be seen as such.
\begin{quote}
Reasoning of this fact can be seen easily by noting 
that the inner product of link variables is invariant
under the constant gauge transformation given by $e^{\theta x +\theta_\mu/2}$ 
i.e., the gauge transformation at the midpoint, and then 
there appear $L$-periodicity in the equation, and the subtraction 
can be inverted to the other side of the inner product with a minus sign.
Explicit proof of this fact goes as follows.
\eqnb
\langle A^g_\mu|\partial_\mu \omega'_t \rangle=
\langle (A^h)^{e^{\theta x}}_\mu|\partial_\mu 
(e^{-\theta x}\omega_t e^{\theta x}) \rangle
\eqne
Let $A_{x,\mu}=A(U_{x,\mu})$ denote the gauge field at a link 
$(x,x+\mu)$.
Then under the gauge transformation of Bloch wave type, $e^{\theta x}$ ,
it holds at each link that 
\eqnb
A^{e^{\theta x}}_{x,\mu}=A(U^{e^{\theta x}}_{x,\mu})
=A(e^{-(\theta x+\theta_\mu/2)}
e^{+\theta_\mu/2}U_{x,\mu}e^{+\theta_\mu/2}e^{\theta x+\theta_\mu/2})
\eqne
Then corresponding to the situation of constant gauge transformation,
it reads that
\eqnb
A(e^{-(\theta x+\theta_\mu/2)}
e^{+\theta_\mu/2}U_{x,\mu}e^{+\theta_\mu/2}e^{\theta x+\theta_\mu/2})
=e^{-(\theta x+\theta_\mu/2)}
A(U^{\theta_\mu}_{x,\mu})
e^{\theta x+\theta_\mu/2}
\eqne
where $U^{\theta_\mu}_{x,\mu}=e^{+\theta_\mu/2}U_{x,\mu}e^{+\theta_\mu/2}$.
And it holds that
\eqab
\partial_\mu (e^{-\theta x}\omega_t e^{\theta x})=
e^{-(\theta x  + \theta_\mu/2)}&&\nonumber\\
(e^{-\theta_\mu /2}\omega_{t,+\mu} e^{\theta_\mu /2}&-&e^{\theta_\mu /2}
\omega_t e^{-\theta_\mu /2})
e^{\theta x+\theta_\mu/2},
\eqae
where $\omega_{t}=\omega_{t}(x)$ and $\omega_{t,+\mu}=\omega_{t}(x+\mu)$.
Thus we obtain that
\eqab
\langle (A^h)^{e^{\theta x}}_\mu|\partial_\mu 
(e^{-\theta x}\omega_t e^{\theta x}) \rangle=&&\nonumber\\
\langle e^{\theta_\mu/2}(A^h)^{\theta_\mu}_\mu e^{-\theta_\mu/2}| 
\omega_{t,+\mu} \rangle
&-&
\langle e^{-\theta_\mu/2}(A^h)^{\theta_\mu}_\mu e^{\theta_\mu/2}| 
\omega_{t} \rangle
\eqae
where $(A^h)^{\theta_\mu}_{x,\mu}=A((U^h)^{\theta_\mu}_{x,\mu})$ and 
it is to be noted that gauge fields appearing in the inner proucts
are $L$-periodic. Thus we can shift safely the expression as
\eqnb
\langle e^{\theta_\mu/2}(A^h)^{\theta_\mu}_\mu e^{-\theta_\mu/2}| 
\omega_{t,+\mu} \rangle
=
\langle e^{\theta_\mu/2}(A^h)^{\theta_\mu}_{\mu,-\mu} e^{-\theta_\mu/2}| 
\omega_{t} \rangle
\eqne
where 
\eqab
(e^{\theta_\mu/2}(A^h)^{\theta_\mu}_{\mu,-\mu} e^{-\theta_\mu/2})_{x,\mu}
&&=A(e^{\theta_\mu}U^h_{x-\mu,\mu})\nonumber\\
&=&e^{\theta x}A(e^{-(\theta x -\theta_\mu)}U^h_{x-\mu,\mu}e^{\theta x})
e^{-\theta x}\nonumber\\
&=&e^{\theta x}A(U^{he^{\theta x}}_{x-\mu,\mu})e^{-\theta x}.
\eqae
Thus it holds that
\eqab
\langle e^{\theta_\mu/2}(A^h)^{\theta_\mu}_\mu e^{-\theta_\mu/2}| 
\omega_{t,+\mu} \rangle
&=&\langle e^{\theta x}A^g_{\mu,-\mu}e^{-\theta x}|\omega_t \rangle\nonumber\\
&=&\langle A^g_{\mu,-\mu}|\omega'_t \rangle
\eqae
Similarly it holds that
\eqnb
\langle e^{-\theta_\mu/2}(A^h)^{\theta_\mu}_\mu e^{\theta_\mu/2}| 
\omega_{t} \rangle
=\langle A^g_{\mu}|\omega'_t \rangle
\eqne
and thus we obtain that
\eqnb
\langle A^g_\mu|\partial_\mu \omega'_t \rangle=
-\langle\partial_\mu A^g_\mu|\omega'_t \rangle
\eqne
\end{quote}
\noindent If we proceed further to a higher derivative in general, then
we obtain as
\eqnb
{d^2 F_U(g(t))\over dt^2}
=2(-\langle \partial_\mu D_\mu(U^g)\xi_t| \omega'_t \rangle
-\langle \partial_\mu A^g_\mu| (\omega'_{t})_t \rangle
+\langle D_\mu(U^g)\xi_t| \eta_\mu \rangle).
\label{FUTT1}
\eqne
Now we consider a situation 
$U^{g(0)}\in \Xi_L$  and $A^{g(0)}_\mu=A_\mu$, and then since
\eqnb
\left .{d\over dt}A^g_\mu\right |_{t=0}=D_\mu(U)(\omega_t+\eta x)
=D_\mu(U)\omega_t + G(U_\mu) \eta_\mu +[A_{x,\mu}, \eta_\mu{\overline{x}}^\mu],
\eqne
the L-periodicity of $A_\mu$ is easily violated due to the third term.
Now we have
\eqnb
\left .{d\over dt}F_U(g(t))\right |_{t=0}=2(-\langle \partial_\mu A_\mu| \omega_t \rangle
+\langle A_\mu|\eta_\mu \rangle),
\label{FUT2}
\eqne
\eqnb
{\left .{d^2 F_U(g(t))\over dt^2}\right |}_{t=0}
=2(-\langle \partial_\mu D_\mu(U)\xi_t| \omega_t \rangle
-\langle \partial_\mu A_\mu| (\omega'_{t})_t \rangle
+\langle D_\mu(U)\xi_t| \eta_\mu \rangle)
\label{FUTT2}
\eqne
where $\xi_t=\omega_t+\eta x$, and then noting $\partial A=0$,
\eqnb
\left .{d\over dt}F_U(g(t))\right |_{t=0}=2\langle A_\mu|\eta_\mu \rangle,
\label{FUT3}
\eqne
\eqnb
\left .{d^2 F_U(g(t))\over dt^2}\right |_{t=0}
=2(\langle  \omega_t | -\partial_\mu D_\mu(U)\xi_t \rangle
+\langle \eta_\mu |D_\mu(U)\xi_t \rangle).
\label{FUTT3}
\eqne
It should follow from the fact $U_\mu\in \Lambda_\infty$ that
\eqnb
\left .{d\over dt}F_U(g(t))\right |_{t=0}=0,
\label{NC1}
\eqne
\eqnb
\left .{d^2 F_U(g(t))\over dt^2}\right |_{t=0}
\ge 0,
\label{NC2}
\eqne
for $^\forall h\in G_L$ and $^\forall \eta_\mu$.
For the estimate, 
\eqnb
\langle \eta_\mu |D_\mu (U)\xi_t \rangle =
\langle \eta_\mu |D_\mu (U)\omega_t + G(U_\mu) \eta_\mu 
+[A_{x,\mu},{\overline{\eta x}}^\mu]\rangle,
\eqne 
we have
\eqab
\langle \eta_\mu |D_\mu (U)\xi_t \rangle 
&=&
\langle \eta_\mu |D_\mu (U)\omega_t\rangle + 
\langle \eta_\mu|G(U_\mu) \eta_\mu \rangle\nonumber\\
&=&
-\langle D_\mu (U) \eta_\mu |\omega_t\rangle + 
\langle \eta_\mu|G(U_\mu) \eta_\mu \rangle
,
\eqae 
where 
\eqnb
\langle \eta_\mu|[A_{x,\mu}, {\overline{\eta x}}^\mu]\rangle
=
\langle A_{x,\mu}|[{\overline{\eta x}}^\mu,\eta_\mu]\rangle
=0
\eqne
is used. From derivation,
\eqab
\partial_\mu D_\mu (U)\xi_t 
&=&
\partial_\mu( D_\mu (U)\omega_t + G(U_\mu) \eta_\mu 
+[A_{x,\mu}, {\overline{\eta x}}^\mu])\nonumber\\
&=&
\partial_\mu D_\mu (U)\omega_t + \partial_\mu G(U_\mu) \eta_\mu 
+[\partial_\mu A_{\mu}, \eta x]
+[\overline{A_\mu}^\mu,\eta_\mu]\nonumber\\
&=&
\partial_\mu D_\mu (U)\omega_t + \partial_\mu G(U_\mu) \eta_\mu 
+\overline{[A_\mu,\eta_\mu]}^\mu\nonumber\\
&=&
\partial_\mu D_\mu (U)\omega_t + D_\mu (U) \eta_\mu ,
\eqae 
we obtain for (\ref{FUTT3}) that
\eqab
{1\over 2}
\left .{d^2 F_U(g(t))\over dt^2}\right |_{t=0}
&=&\langle  \omega_t | -\partial_\mu D_\mu(U)\omega_t \rangle
-\langle  \omega_t | D_\mu (U) \eta_\mu\rangle\nonumber\\
&&-\langle D_\mu (U) \eta_\mu |\omega_t\rangle + 
\langle \eta_\mu|G(U_\mu) \eta_\mu \rangle,
\eqae
\eqab
{1\over 2}
\left .{d^2 F_U(g(t))\over dt^2}\right |_{t=0}
&=&\langle  \omega_t-\dis{1\over -\partial D}D\eta | -\partial D|
\omega_t -\dis{1\over -\partial D}D\eta\rangle\nonumber\\
&&-\langle  D\eta|\dis{1\over -\partial D}|D\eta\rangle
+\langle \eta_\mu|G(U_\mu)| \eta_\mu \rangle
\label{FUTT4}
\eqae
Now we draw some necessary conditions for $U_\mu\in \Xi_L$ from 
(\ref{FUT3}), (\ref{NC1}), (\ref{NC2}) and (\ref{FUTT4}).\\

{\bf Theorem 2} \\
Let $U_\mu\in \Xi_L=\Lambda_{\infty}\cap \Pi_L$. 
Then it follows that for the gauge transformation 
$g=e^{\omega t}e^{\eta x t}$ where $^\forall \omega$ belongs 
to $L$-periodic scalar
and $^\forall \eta_\mu$ belongs to the same Cartan subalgebra, 
the optimizing fucntion
$ F_U(g)$ behaves as
\eqnb
\left .{d\over dt}F_U(g(t))\right |_{t=0}=2\langle A_\mu|\eta_\mu \rangle=0,
\label{FUT5}
\eqne
and
\eqab
{1\over 2}
\left .{d^2 F_U(g(t))\over dt^2}\right |_{t=0}
&=&\langle  \omega -\dis{1\over -\partial D}D\eta | -\partial D|
\omega -\dis{1\over -\partial D}D\eta\rangle\nonumber\\
&&-\langle  D\eta|\dis{1\over -\partial D}|D\eta\rangle
+\langle \eta_\mu|G(U_\mu)| \eta_\mu \rangle\ge 0.
\label{FUTT5}
\eqae
Thus it is concluded that
\eqnb
\bar A_\mu\equiv\sum_x A_{x,\mu}=0
\eqne
and
\eqnb
\langle  D\eta|\dis{1\over -\partial D}|D\eta\rangle
-\langle \eta_\mu|G(U_\mu)| \eta_\mu \rangle\le 0.
\eqne

{\bf Proof} is selfevident from the fact that since $D\eta$ is $L$-periodic for 
$^\forall \eta$, one can choose $\omega$ such that
\eqnb
\omega -\dis{1\over -\partial D}D\eta=0.
\eqne

\section {Horizon function, horizon condition and Kugo-Ojima color confinement criterion}
{\bf Horizon tensor} is defined as
\eqnb
H_{\mu\nu}=-D_\mu(-\partial D)^{-1}D_\nu
-\delta_{\mu\nu}G(U_\mu).
\eqne
Taking the trace of the operator $H_{\mu\nu}$ with respect to the normalized
constant colored vectors $\eta_\mu^{\nu,a}=\delta_{\mu\nu}\lambda_a$ with 
${\rm tr}\lambda_a^{\dag}\lambda_b=\delta_{ab}$, 
one defines the {\bf horizon function} $H(U)$ as 
\eqab
H(U)&&=\sum_{\nu,a}\langle \eta_\mu^{\nu,a}|H_{\mu\rho}|
\eta_\rho^{\nu,a}\rangle\nonumber\\
&&=
\sum_{\nu,a}\langle \eta_\mu^{\nu,a}|
-D_\mu(-\partial D)^{-1}D_\rho
|\eta_\rho^{\nu,a}\rangle
-(N^2-1)E(U)\nonumber\\
&&\equiv h(U)V\ 
\label{HRZNFNCT}
\eqae
where
\eqnb
(N^2-1)E(U)=\sum_{x,\mu,a}{\rm tr}(\lambda_a{\dag} G(U_{x,\mu})\lambda_a).
\eqne 
From Theorem 2, one has for $U\in \Xi_L$ that 
\eqnb
\overline {A_\mu}=V^{-1}
\sum_x A_{x,\mu}=0
\eqne
and 
\eqnb
H(U)\le 0,
\eqne
where $V=L^4$.

Zwanziger hypothesized that {\em the dynamics on $\Xi_L$ tends to
that on $\Lambda_L$ in the infinite volume limit}, and derived 
the {\bf horizon condition}, statistical 
average 
\eqnb
\bigl \langle h(U) \bigr \rangle=0,
\label{HRZNCND}
\eqne
in the infinite volume limit.
Taking the Fourier transform of the tensor propagator of the color point source, 
\eqnb
\bigl \langle -D_\mu(-\partial D)^{-1}D_\nu \bigr \rangle_{xa,yb},
\eqne
one has, with assumption of the global color symmetry not broken, 
\eqnb
G_{\mu\nu}(p)\delta^{ab}=\delta^{ab}[
(e/d)(p_\mu p_\nu/ p^2)
-\{\delta_{\mu\nu}-(p_\mu p_\nu/ p^2)\}u(p^2)],
\label{TNSR}
\eqne
where 
\eqnb
e=\langle E(U)\rangle/V
\eqne
and dimension $d=4$.\par

In local operator formalism of QCD, Kugo and Ojima proposed color confinement criterion\cite{rf:3} based on the BRST(Becchi-Rouet-Stora-Tyutin) symmetry without Gribov's problem taken into account.
Kugo-Ojima two-point function in the continuum theory is defined in the lattice Landau gauge QCD as
\eqnb
(\delta_{\mu\nu}-{p_\mu p_\nu\over p^2})u^{ab}(p^2)=
\dis{1\over V}
\sum_{x,y} e^{-ip(x-y)} \left \langle  {\rm tr}({\lambda^a}^{\dag}
D_\mu \dis{1\over -\partial D}[A_\nu,\lambda^b] )_{xy}
\right \rangle
\eqne
where $u^{ab}(p^2)=\delta^{ab}u(p^2)$ and it was shown that the sufficient condition of color confinement is given by 
\eqnb
u(0)=-1. 
\eqne
Putting Kugo-Ojima parameter as 
\eqnb
u(0)=-c
\eqne
and comparing 
\eqnb
\lim_{p_\mu\to +0}
G_{\mu\mu}(p)
\label{MOMLMT}
\eqne
with
\eqnb
\bigl \langle h(U) \bigr \rangle=0
\eqne
one finds that the horizon condition 
reduces to
\eqnb
\left \langle {h(U)\over N^2-1}\right \rangle
=\left(\dis{e\over d}\right)+(d-1)c-e=(d-1)\left(c-\dis{e\over d}\right)
\equiv (d-1)h=0.
\eqne
Kugo-Ojima's and Zwanziger's arguments emerge to be consistent
with each other provided the lattice covariant derivative naturally meets with 
the continuum one 
\eqnb
e/d=1.
\eqne
This fact was pointed out and some numerical data were presented.\cite{rf:6,rf:7}
\section{Horizon condition in case of free boudary condition}
Let us consider the following quantity
\eqnb
\langle \partial_\mu \phi|D_\mu\phi\rangle
\eqne
on the arbitrary networks with assumptions that $\partial A=0$ in the sense of 
(\ref{DIVLESS}), i.e.,
\eqnb
\partial_\mu A_\mu=\sum_{\mu\in \ell_+(x)}(A_\mu)_+ -
\sum_{\mu\in \ell_-(x)}(A_\mu)_- 
\eqne
where we put indices $_{\pm}$ for link-field $A_\mu$ if $\mu\in \ell_{\pm}(x)$,
and with an assumption that a scalar function $\phi$ does not have 
the zero-eigenvalue eigenvector component of $-\partial D(U)$. 
It follows that
\eqab
\langle \partial_\mu \phi|D_\mu\phi\rangle&=&
\langle \phi| -\partial D|\phi\rangle\nonumber\\
&=&\langle \phi|(-\partial D)\dis{1\over -\partial D}|(-\partial D)\phi\rangle
\nonumber\\
&=&\langle -D\partial\phi|\dis{1\over -\partial D}|(-\partial D)\phi\rangle.
\label{BDID1}
\eqae
One finds the Faddeev-Popov operator is symmetric when $\partial A=0$ which
is seen below.
\eqnb
D_\mu \partial_\mu \phi=\partial_\mu G(U_\mu)\partial_\mu \phi +
\overline{[A_\mu,\partial_\mu \phi]}^\mu,
\eqne
\eqnb
\partial_\mu D_\mu \phi=\partial_\mu G(U_\mu)\partial_\mu \phi +
\partial_\mu [A_\mu,{\bar\phi}^\mu],
\eqne
\eqab
\left( \overline{[A_\mu,\partial_\mu \phi]}^\mu
-\partial_\mu [A_\mu,{\bar\phi}^\mu]\right )_x
&=&\sum_{\mu\in \ell_+(x)}
\left([A_{\mu},\dis{1\over 2}(\partial_\mu\phi)]_+ -[A_{\mu},{\bar\phi}^\mu]_+
\right)\nonumber\\
&&+
\sum_{\mu\in \ell_-(x)}
\left([A_{\mu},\dis{1\over 2}(\partial_\mu\phi)]_- +[A_{\mu},{\bar\phi}^\mu]_-
\right)\nonumber\\
&=&-\left (\sum_{\mu\in \ell_+(x)}
[A_{\mu},{\bar\phi}^\mu-\dis{1\over 2}(\partial_\mu\phi)]_+\right .\nonumber\\
&&+
\left .\sum_{\mu\in \ell_-(x)}
[-A_{\mu},{\bar\phi}^\mu+\dis{1\over 2}(\partial_\mu\phi)]_-
\right)\nonumber\\
&=&-\left [\sum_{\mu\in \ell_+(x)}A_{\mu}-\sum_{\mu\in \ell_-(x)}A_{\mu},\ 
\phi_x\right ]\nonumber\\
&=&-[\partial_\mu A_{\mu},\phi_x ].
\eqae
Thus it hold that if $\partial A=0$, then
\eqnb
D_\mu \partial_\mu \phi=\partial_\mu D_\mu \phi,
\eqne
and
\eqnb
\langle D\partial\phi|\dis{1\over -\partial D}|D\partial\phi\rangle
-\langle \partial_\mu \phi|D_\mu\phi\rangle=0.
\label{HZID}
\eqne
This equation (\ref{HZID}) is {\em an identity which holds on any networks 
when $\partial A=0$ and $\phi$ is free from zero-eigenvalue 
eigenvector component of $-\partial D$}.

Now let us consider the case with free boundary condition of regular lattice 
$L^d$. and let us assume 
$\phi=x\eta$ where $\eta_\mu$'s are suitably normalized antihermitian matrices 
such that
\eqnb
[\eta_\mu,\eta_\nu]=0.
\eqne
Then it holds from 
\eqnb
\partial_\mu \phi=\eta_\mu
\label{RTSNLSS}
\eqne
and from 
\eqnb
\langle \eta_\mu|[A_\mu,\overline{\eta x}^\mu]\rangle
=\langle A_\mu|[\overline{\eta x}^\mu,\eta_\mu]\rangle
=0
\eqne
that
\eqnb
\langle D_\mu \eta_\mu|\dis{1\over -\partial D}|D_\nu\eta_\nu\rangle
-\langle \eta_\mu|G(U_\mu)|\eta_\mu\rangle=0.
\label{HZFNC1}
\eqne
By putting
\eqnb
\eta^{\rho,a}_\mu=\delta_{\mu\rho}\lambda_a,
\label{HZFNC1_1}
\eqne
one obtains 
\eqnb
D_\mu\eta^{\rho,a}_\mu=D_\nu\eta^{\rho,a}_\nu=D_\rho \lambda_a
\eqne
and then the vanishing horizon function
\eqnb
H(U)=\sum_{\rho,a}\left (\langle D_\rho \lambda_a|\dis{1\over -\partial D}|
D_\rho\lambda_a\rangle
-\langle \lambda_a|G(U_\rho)|\lambda_a\rangle\right )=0.
\label{HZFNC2}
\eqne
It is to be noted that $\lambda_a$ in equation (\ref{HZFNC2}) is located on links, 
and as seen from (\ref{HZFNC1}), $D_\rho$ acts on $\lambda_a$ as defined in (\ref{CVDR2}) 
with non-vanishing first term.
\section{Discussions and conclusions}
In \S1, notations and generalities are given in order to discuss the case of free boundary 
condition from more general point of view, i.e., network QCD. The main purposes are to 
discuss behavior of the optimizing function for Landau gauge, gauge non-invariant function, 
and we need extended definitions, e.g., $\partial_\mu \phi$ and $\partial_\mu B_\mu$, with 
clear distinction between site-functions and link-functions, and finally to obtain an  identity which holds on arbitrary networks. As a matter of course, full formulation of 
network QCD is out of scope of the present study.

In \S2, definitions of gauge field, $U$-liner type and $\log U$-type, are given, together with covariant derivatives for 
each type, respectively, where difference between covariant derivative (\ref{CVDR1}) and covariant divergence (\ref{CVDR2}) should be noted.

In \S3, following Ref.~\citen{rf:2}, various kinds of regions in Landau gauge are 
defined, i.e., Gribov region, $\Omega_L$, fundamental modular region, $\Lambda_L$, and 
core region, $\Xi_L$, on regular lattice of period $L$, where the following inclusions 
hold,
\eqnb
\Xi_L=(\Lambda_{\infty}\cap \Pi_L) \subset \Lambda_L \subset \Omega_L.
\eqne
Theorem 2 states that for each $U_\mu\in \Xi_L=\Lambda_{\infty}\cap \Pi_L$, the horizon 
function defined in \S4 (\ref{HRZNFNCT}) takes non-positive value, $H(U)\le 0$.

In \S4, although we have skipped the derivation of the horizon condition, (\ref{HRZNCND}), 
Zwanziger showed in Ref.~\citen{rf:2} that it can be derived from statistical average on 
augmented core region, $\Psi_L$, in infinite volume limit, $L\to \infty$, where
\eqnb
\Xi_L \subset \Psi_L\equiv \{U|\ H(U)\le 0,\ U\in \Omega_L\}\subset \Omega_L.
\eqne 
It is reviewed that the horizon condition and the Kugo-Ojima criterion of the color confinement\cite{rf:3} coincide with each other in the continuum limit\cite{rf:6,rf:7}.

In \S5, we focuss on the fact that the horizon condition holds for each configuration in Landau gauge on finite regular lattice with the free boundary condition.\cite{rf:5}
It is found that the fact can be derived from the equation (\ref{HZID}),
\[
\langle D\partial\phi|\dis{1\over -\partial D}|D\partial\phi\rangle
-\langle \partial_\mu \phi|D_\mu\phi\rangle=0,
\]
that is {\em an identity which holds on any networks when $\partial A=0$ and $\phi$ is free from zero-eigenvalue eigenvector component of $-\partial D$}.
Obvious reason why vanishing horizon function (\ref{HZFNC2}) is 
not realized in the case 
of the periodic boundary condition is that (\ref{RTSNLSS}) does not hold.
In a special case of $U=1$ with free boundary condition, (\ref{HZFNC2}) turns out to be
\eqnb
H(U)=\sum_{\rho,a}\left (\langle \partial_\rho \lambda_a|
\dis{1\over -\partial^2}|
\partial_\rho\lambda_a\rangle
-\langle \lambda_a|1|\lambda_a\rangle\right )=0,
\label{HZFNC3}
\eqne
where there appear $\pm\delta_x$ 'charge density' only on the boundary surfaces.
However, in the generic non-constant $U$ cases, 'charge density' $D_\rho\lambda_a$ 
in (\ref{HZFNC2}) spreads over $d$-dimensional volume. Thus one of the most important open 
questions is if the boundary condition affects the physics in the 4-dimensional bulk 
system in the thermodynamic limit, i.e., continuum limit.

\section*{Acknowledgments} 
The author would like to express sincere gratitude to Professor Daniel Zwanziger for 
reading the draft with great interest some time ago, and for giving him encouraging comments then. He would also like to thank Taichiro Kugo for reading the draft and 
encouraging him to publish it. He appreciates enlightening discussion on horizon functions with Taichiro Kugo, Kei-ichi Kondo and Hideo Suganuma.


\end{document}